%% file: sample-sigconf.tex
\documentclass[sigconf]{acmart}

\usepackage{graphicx}
\usepackage{subcaption}

\usepackage{booktabs}
\usepackage{siunitx}
\AtBeginDocument{%
  }

\setcopyright{none}
\copyrightyear{2018}
\acmYear{}

\copyrightyear{2025}
\acmYear{2025}
\setcopyright{cc}
\setcctype{by}
\acmConference[SC Workshops '25]{Workshops of the International Conference for High Performance Computing, Networking, Storage and Analysis}{November 16--21, 2025}{St Louis, MO, USA}
\acmBooktitle{Workshops of the International Conference for High Performance Computing, Networking, Storage and Analysis (SC Workshops '25), November 16--21, 2025, St Louis, MO, USA}
\acmDOI{10.1145/3731599.3767522}
\acmISBN{979-8-4007-1871-7/2025/11}

\begin{document}

\title{An RDMA-First Object Storage System with SmartNIC Offload}

\author{Yu Zhu}
\affiliation{%
  \institution{ETH Zurich}
  \country{Switzerland}
}
\email{yu.zhu@inf.ethz.ch}

\author{Aditya Dhakal}
\affiliation{%
  \institution{Hewlett Packard Labs}
  \country{United States}
  }
\email{aditya.dhakal@hpe.com}
\author{Pedro Bruel}
\affiliation{%
  \institution{Hewlett Packard Labs}
  \country{United States}
}
\email{bruel@hpe.com}
\author{Gourav Rattihalli}
\affiliation{%
  \institution{Hewlett Packard Labs}
  \country{United States}
}
\email{gourav.rattihalli@hpe.com}
\author{Yunming Xiao}
\affiliation{%
    \institution{\hspace*{-1.2em}\mbox{Chinese University of Hong Kong, Shenzhen}}
    \country{China}
}
\email{yunmingxiao@cuhk.edu.cn}
\author{Johann Lombardi}
\affiliation{%
  \institution{Hewlett Packard Enterprise}
  \country{France}
}
\email{johann.lombardi@hpe.com}

\author{Dejan Milojicic}
\affiliation{%
  \institution{Hewlett Packard Labs}
  \country{United States}
}
\email{dejan.milojicic@hpe.com}

\renewcommand{\shortauthors}{Zhu et al.}


\begin{abstract}
AI training and inference impose sustained, fine-grain I/O that stresses host-mediated, TCP-based storage paths. 
Motivated by kernel-bypass networking and user-space storage stacks, we revisit POSIX-compatible object storage for GPU-centric pipelines. 
We present \emph{ROS2}, an RDMA-first object storage system design that offloads the DAOS client to an NVIDIA BlueField-3 SmartNIC while leaving the DAOS I/O engine unchanged on the storage server. 
ROS2 separates a lightweight control plane (gRPC for namespace and capability exchange) from a high-throughput data plane (UCX/libfabric over RDMA or TCP) and removes host mediation from the data path.

Using FIO/DFS across local and remote configurations, we find that on server-grade CPUs RDMA consistently outperforms TCP for both large sequential and small random I/O. 
When the RDMA-driven DAOS client is offloaded to BlueField-3, end-to-end performance is comparable to the host, demonstrating that SmartNIC offload preserves RDMA efficiency while enabling DPU-resident features such as multi-tenant isolation and inline services (e.g., encryption/decryption) close to the NIC. 
In contrast, TCP on the SmartNIC lags host performance, underscoring the importance of RDMA for offloaded deployments.

Overall, our results indicate that an RDMA-first, SmartNIC-offloaded object-storage stack is a practical foundation for scaling data delivery in modern LLM training environments; integrating optional GPU-direct placement for LLM tasks is left for future work.
\end{abstract}



\maketitle

\input{introduction}

\input{background}

\input{implementation}

\input{evaluation}

\input{discussion}

\input{related_work}

\input{conclusion}

\bibliographystyle{ACM-Reference-Format}  
\bibliography{sample-base}

\end{document}

%% file: introduction.tex
\section{Introduction}
Remote Direct Memory Access (RDMA) has emerged as a critical technology for high-performance data movement,
enabling low-latency and high-throughput communication by bypassing the CPU in the data path and allowing direct memory-to-memory transfers between endpoints \cite{mittal2018revisiting, gao2021cloud}.
While RDMA has been widely adopted in traditional high-performance computing (HPC) environments,
the recent prosperity of artificial intelligence (AI) workloads, especially large language models (LLM), introduces a new set of requirements that differ fundamentally from those of conventional HPC tasks.
Modern AI training pipelines \cite{gangidi2024rdma, zhao2025insights}, particularly those involving large-scale models, demand frequent access to massive datasets with stringent latency and throughput constraints,
making efficient data movement an indispensable factor for end-to-end performance.

\begin{figure}[h]
  \centering
  \includegraphics[width=\linewidth]{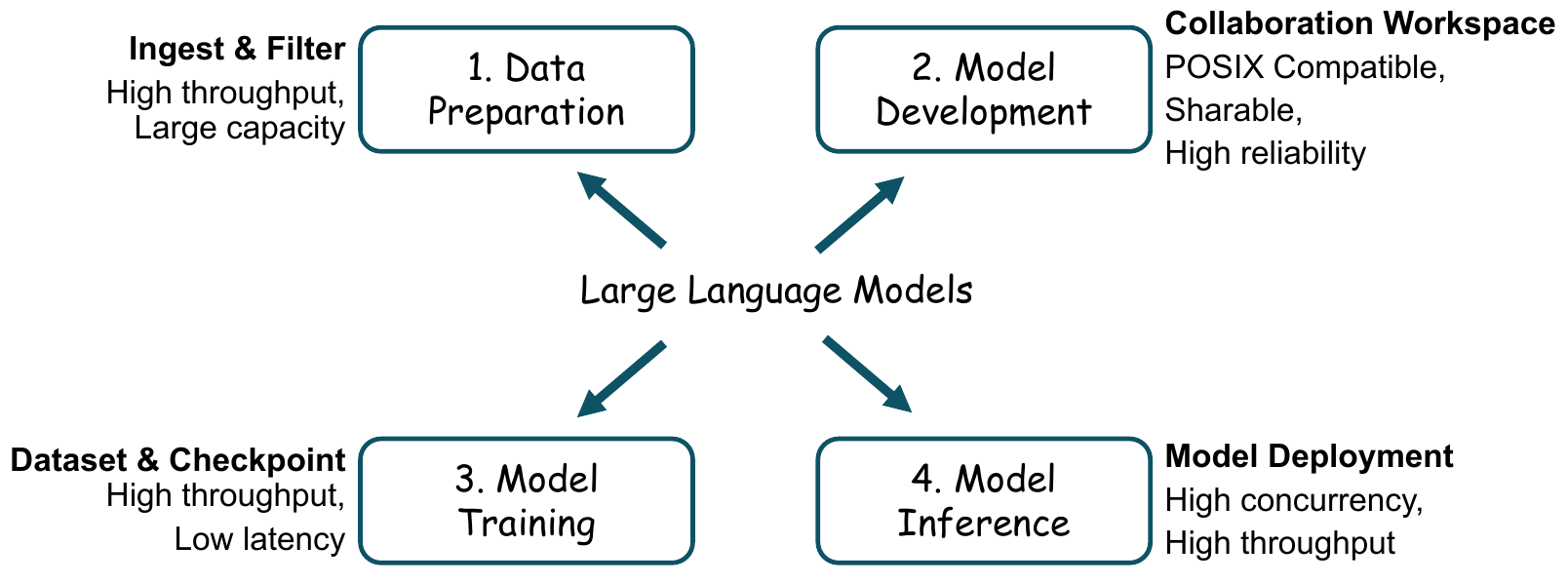}
  \caption{Diverse requirements of storage for LLM tasks.}
  \label{fig:storage_requirement}
\end{figure}

Despite RDMA's proven advantages, it remains largely absent from contemporary cloud object storage services,
which rely on TCP-based protocols over HTTP for their accessibility, scalability, and multi-tenancy \cite{mazumdar2019survey, mondal2024demystifying}.
These designs, while robust for general-purpose workloads, introduce significant software stack overheads and network inefficiencies
that become bottlenecks for AI training at scale.
In contrast, recent efforts such as DeepSeek 3FS \cite{deepseek3fs} have demonstrated the substantial performance benefits of applying RDMA to distributed storage in AI training environments,
achieving dramatic reductions in data access latency and improved I/O parallelism.

\begin{table*}[t]
\centering
\caption{NVIDIA data center GPUs across generations (representative configurations).}
\label{tab:nvidia-gpus}
\begin{tabular}{l l l l l l l l}
\toprule
GPU & Architecture & Memory (GB) & Mem BW & NVLink (gen / per-GPU BW) & FP16 & FP8 & FP4 \\
\midrule
P100 \cite{nvidia_p100} & Pascal   & 16 HBM2   & 732~GB/s                & NVLink 1 / up to 80~GB/s   & 21.2 TFLOPS & N/A        & N/A \\
V100 \cite{nvidia_v100} & Volta    & 32 HBM2   & 1134~GB/s               & NVLink 2 / up to 300~GB/s  & 130 TFLOPS\textsuperscript{\dag} & N/A & N/A \\
A100 \cite{nvidia_a100} & Ampere   & 80 HBM2e  & $\approx$2.0~TB/s       & NVLink 3 / up to 600~GB/s  & 624 TFLOPS  & N/A        & N/A \\
H100 \cite{nvidia_h100} & Hopper   & 80 HBM3   & 3.35~TB/s               & NVLink 4 / up to 900~GB/s  & $\approx$2 PFLOPS & $\approx$4 PFLOPS & N/A \\
H200 \cite{nvidia_h200} & Hopper   & 141 HBM3e & 4.8~TB/s                & NVLink 4 / up to 900~GB/s  & $\approx$2 PFLOPS & $\approx$4 PFLOPS & N/A \\
B200 \cite{nvidia_b200} & Blackwell& 186 HBM3e & 8.0~TB/s                & NVLink 5 / up to 1.8~TB/s  & 5 PFLOPS    & 10 PFLOPS  & 20 PFLOPS \\
\bottomrule
\end{tabular}

\vspace{2pt}
\begin{minipage}{0.95\linewidth}
\footnotesize \textsuperscript{\dag}\,V100 FP16 figure is \emph{Tensor Core} throughput (FP16 inputs, FP32 accumulate).  
Scalar FP16 on CUDA cores is $\sim$28–33 TFLOPS depending on variant and clocks.
\end{minipage}
\end{table*}


In this work we integrate RDMA into an object-storage stack and quantify its advantages over TCP for AI data paths. Using Distributed Asynchronous Object Storage (DAOS \cite{hennecke2020daos}) as a representative system, we offload the POSIX-compatible DFS (DAOS File System) client/data plane to an NVIDIA BlueField-3 SmartNIC \cite{bluefield-3} to reduce host involvement and enable stronger multi-tenant isolation. 
Our evaluation measures end-to-end throughput and IOPS for remote NVMe access under TCP and RDMA on server-grade CPUs and a BlueField-3 SmartNIC. In our experiments, RDMA generally outperformed TCP on the host, and RDMA on the SmartNIC achieved performance comparable to the host, whereas SmartNIC-based TCP was markedly lower. We also outline an optional extension for direct GPU placement via GPUDirect RDMA \cite{gpudirectrdma}; implementing and evaluating this extension is beyond the scope of this paper and part of ongoing work.


With this paper, we make the following contributions:
\begin{itemize}
  \item We implement an \emph{RDMA-first}, POSIX-compatible DAOS client on an NVIDIA BlueField-3 SmartNIC, separating a small gRPC control plane from a high-throughput data plane (UCX \cite{ucx} or libfabric \cite{libfabric} over RDMA or TCP) and keeping the host CPU off the fast path.
  \item We conduct a systematic evaluation spanning raw NVMe baselines, remote SPDK, and end-to-end DAOS/DFS on server-grade CPUs and BlueField-3.
  \item Our prototype sustains near line-rate throughput for large sequential transfers and demonstrates that SmartNIC offload is a practical way to reduce host involvement while preserving performance; we outline optional direct GPU placement via GPUDirect RDMA as future work.
\end{itemize}

%% file: background.tex
\section{Background}

\subsection{Fast Evolution of GPUs}
Table~\ref{tab:nvidia-gpus} summarizes NVIDIA data-center GPUs from Pascal (P100) to Blackwell (B200) using representative server configurations. 
This evolution is characterized by three architectural trends. 
\emph{First}, on-package memory has expanded substantially—from 16\,GB HBM2 at 0.73\,TB/s (P100) to roughly 180\,GB HBM3e at 8\,TB/s (B200)—thereby increasing the sustained data rate that storage subsystems must deliver. 
\emph{Second}, inter-GPU interconnect bandwidth has advanced from NVLink~1 to NVLink~5 (up to $\sim$1.8\,TB/s per GPU), making intra-node exchange markedly faster than host-mediated I/O paths. 
\emph{Third}, compute throughput has progressed from scalar FP16 to Tensor-Core precisions (FP16/BF16, FP8, and FP4), moving peak tensor performance from TFLOPS to multi-PFLOPS per GPU.

\textbf{Implications for LLM data ingestion.}
As HBM bandwidth and tensor throughput rise, LLM training becomes increasingly sensitive to the efficiency of data delivery. 
Let
\[
B_{\text{node}} \approx G \cdot r \cdot s
\]
denote the required sustained ingest rate per node, where $G$ is the number of GPUs, $r$ is the per-GPU sample/token rate, and $s$ is the average bytes fetched per sample after compression. 
Even conservative choices yield multi-GiB/s per node and substantial small-I/O pressure due to shuffling and mixed datasets. 
Under these conditions, host-mediated TCP paths incur nontrivial per-I/O overhead and CPU utilization, whereas RDMA’s kernel-bypass, zero-copy transfers better match GPU/NVLink speeds. 
These trends motivate the RDMA-first, SmartNIC-offloaded object-storage design evaluated in this work.

\subsection{DeepSeek 3FS}

DeepSeek 3FS \cite{deepseek3fs}, also known as the \textit{Fire-Flyer File System}, is a distributed storage system
explicitly designed to meet the extreme I/O demands of large-scale AI training and inference.
Its architecture tightly integrates NVMe SSD storage with RDMA-capable high-speed networks
to enable direct, low-latency, and high-throughput data access from compute nodes.

\textbf{RDMA-centric design.}
3FS employs RDMA as its primary transport, allowing compute nodes to directly read from and write to
remote NVMe storage without intermediate CPU copies or kernel involvement.
This zero-copy, kernel-bypass communication model significantly reduces data access latency
and maximizes bandwidth utilization, which is critical when scaling AI training to thousands of GPUs.
In production deployments with 200~Gbps InfiniBand and 180 storage servers,
3FS has achieved an aggregate read bandwidth of 6.6~TiB/s.

\textbf{Alignment with AI workflow patterns.}
3FS targets the data-access characteristics of AI pipelines, including (i) large sequential parameter loading, (ii) high-concurrency random reads for data loaders, and (iii) asynchronous checkpointing during training.
Its RDMA-based client–storage path reduces kernel traversal and host copies relative to TCP, and—combined with asynchronous zero-copy API (inspired by Linux \texttt{io\_uring} \cite{io_uring})—exposes enough parallelism to saturate network links and NVMe back ends.
At present, 3FS does not explicitly employ GPU-direct data placement (e.g., GPUDirect RDMA/Storage); incorporating such mechanisms is a natural enhancement to further reduce host mediation and PCIe traffic, but lies outside the scope of current deployments.

\textbf{Relevance to RDMA-based object storage research.}
The success of 3FS highlights the benefits of RDMA in distributed storage for AI workloads—predictable low latency, near–line-rate throughput, and efficient multi-node scaling.
These properties make 3FS a useful point of reference both for RDMA-first systems such as DAOS and for exploring RDMA integration behind or alongside the HTTP/TCP interfaces of public cloud object stores.

\subsection{Security Risks of RDMA in Clouds}

RDMA enables low-latency, zero-copy data transfer by granting peers direct memory access via
\textit{rkeys} issued at memory registration.
Possession of a valid rkey, along with the target address, allows remote read/write
without CPU involvement or higher-level authentication.
While suitable for trusted HPC clusters, this capability model poses risks in public clouds \cite{tsai2019pythia, tsai2019double, simpson2020securing, xing2022bedrock}.

The primary concerns are:
\begin{itemize}
    \item \textbf{Cross-tenant access:} 
    Misconfiguration, overly broad memory regions, or rkey leakage can let one tenant read or corrupt another's data.
    \item \textbf{Bypassing access control:} 
    RDMA operations skip software checks, enabling undetected high-speed data exfiltration once an rkey is compromised.
    \item \textbf{Weak isolation:} 
    Shared protection domains or insufficient IOMMU/SR-IOV enforcement can allow DMA outside assigned memory.
\end{itemize}

Building on concerns raised by \textit{Pythia}~\cite{tsai2019pythia}, exposing low-level RDMA in multi-tenant settings can increase risk.
Offloading the RDMA-driven client stack to a BlueField-3 DPU \cite{bluefield-3} has the potential to narrow the host attack surface and enable finer-grained controls—e.g., per-tenant protection domains/QPs, short-lived scoped \texttt{rkey}s, strict memory registration, and fabric segmentation—while keeping policy enforcement close to the NIC \cite{grant2020smartnic, zhou2024smartnic}.

\subsection{Distributed Asynchronous Object Storage}

The Distributed Asynchronous Object Storage (DAOS \cite{daos, hennecke2023understanding}) system is an open-source, high-performance object store
originally developed by Intel to address the extreme I/O requirements of next-generation HPC and data-intensive workloads.
DAOS departs from traditional disk-based storage architectures by adopting a fully disaggregated design
that exploits persistent memory (PMEM \cite{pmem}) and NVMe solid-state drives (SSDs) as the primary storage media,
achieving orders-of-magnitude lower latency and higher throughput compared to HDD-based systems.

At its core, DAOS implements a transactional, versioned object model.
Objects are distributed across a set of storage targets, each managed by a DAOS server process,
and can store both structured and unstructured data.
DAOS separates metadata and data paths, employing a scalable key–array data layout and end-to-end checksums to ensure consistency and reliability.
Its software stack is built on top of the Storage Performance Development Kit (SPDK \cite{spdk}) for NVMe access
and leverages the Mercury RPC framework for inter-node communication.


DAOS integrates with high-speed RDMA fabrics (e.g., InfiniBand \cite{pfister2001introduction}, RoCEv2 \cite{guo2016rdma, rocev2}) via UCX/libfabric and uses kernel-bypass networking and SPDK. The native libdaos API and higher-level integrations (MPI-IO, HDF5 via the DAOS VOL, and POSIX via DFS/dfuse) allow DAOS to serve as a common back end for HPC simulations and emerging AI/analytics workloads.

By combining user-space networking, persistent memory, and NVMe, DAOS aims to deliver extreme I/O concurrency,
low tail latency, and scalable metadata operations—characteristics that make it a compelling candidate
for exploring RDMA optimizations and SmartNIC offloading in object storage systems.

\subsection{SmartNICs and the NVIDIA BlueField-3}

A SmartNIC \cite{kfoury2024comprehensive} is a network interface card equipped with programmable processing capabilities
that offload networking, storage, and security functions from the host CPU.
By executing these functions directly on the NIC, SmartNICs reduce host CPU utilization,
improve I/O performance, and enable in-network data processing.
They are particularly valuable in data centers and cloud environments
where isolation, performance, and scalability are critical.

The NVIDIA BlueField-3 is a modern SmartNIC that integrates a high-speed ConnectX-7 NIC
with an embedded system-on-chip (SoC) featuring up to 16 Arm Neoverse cores, accelerators
for cryptography and data compression, and hardware support for RDMA over Converged Ethernet (RoCEv2)
and InfiniBand.
It provides secure, isolated execution environments for network and storage services,
including NVMe-over-Fabrics (NVMe-oF \cite{nvme_over_fabric}) and RDMA transport termination,
allowing functionality such as firewalling, encryption, or storage protocol handling
to be performed entirely on the NIC.
These capabilities make BlueField-3 a strong candidate for offloading object storage
data paths, enabling lower latency, improved throughput, and enhanced multi-tenant isolation.

%% file: implementation.tex
\section{Design and Implementation}\label{sec:design}

\begin{figure}[t]
  \centering
  \includegraphics[width=\linewidth]{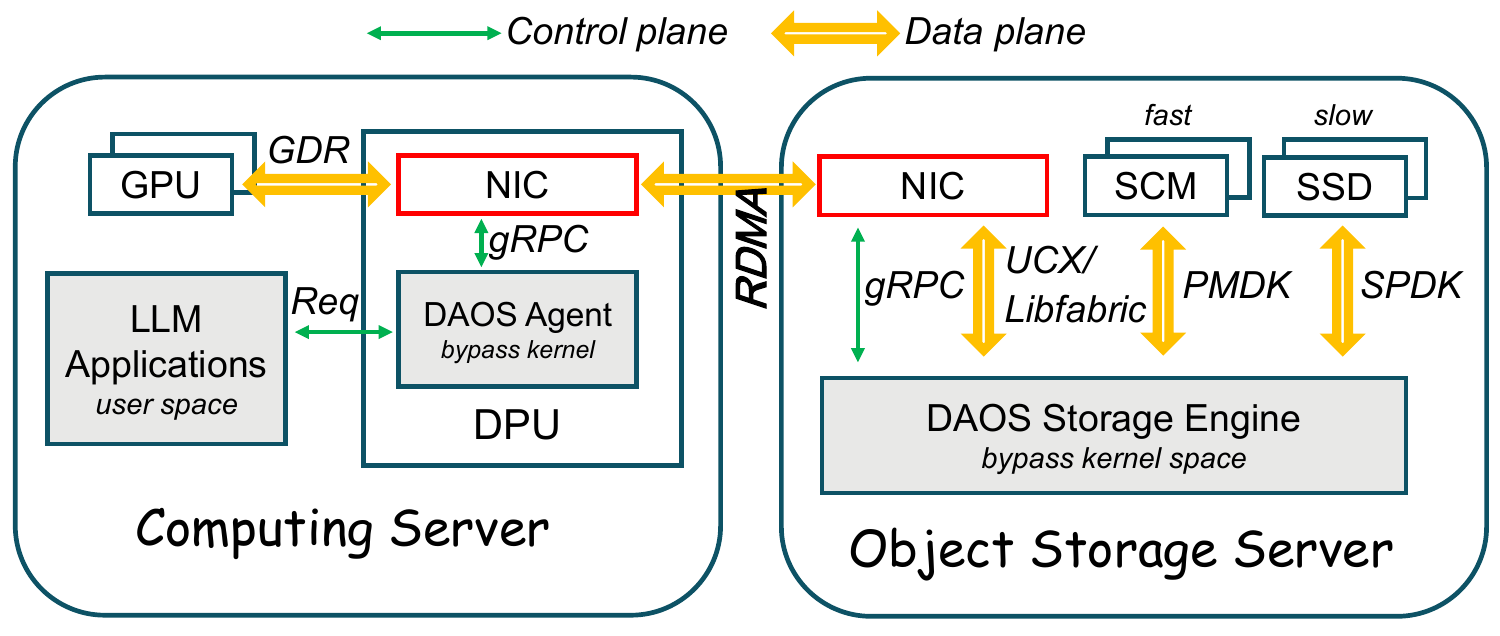}
  \caption{RDMA-first design with SmartNIC-offloaded client and kernel-bypass storage engine.}
  \label{fig:system_design}
\end{figure}

\subsection{Overview}
Figure~\ref{fig:system_design} shows our architecture of ROS2 for POSIX-style file I/O over DAOS.
We target GPU-centric LLM pipelines but evaluate the \emph{data path} with standard POSIX operations (read, write, randread, randwrite) generated by FIO using the DAOS DFS engine.
The key idea is to relocate the DAOS client functionality to a SmartNIC (NVIDIA BlueField-3 DPU) on the computing server, while keeping the DAOS I/O engine unmodified on the object-storage server.
We split the system into a lightweight control plane (green in the figure) and a high-throughput data plane (yellow).
The control plane uses gRPC \cite{grpc} for session setup, authentication, and namespace/DFS metadata operations; the data plane uses user-space networking (UCX or libfabric) over either TCP or RDMA to transfer file data.
On the storage side, DAOS accesses storage tiers entirely in user space: Storage Class Memory (SCM \cite{nider2020processing}) via Persistent Memory Development Kit (PMDK \cite{pmdk}) and NVMe via SPDK.

\subsection{Computing Server: POSIX-on-DPU}
\textbf{DFS client and FIO.}
The DFS client stack (libdaos/libdfs) executes on the DPU.
For benchmarking we run FIO with the DFS engine, which issues POSIX-style file calls (\texttt{open}, \texttt{read}, \texttt{write}, \texttt{fsync}, etc.) that the DFS client translates to DAOS RPCs and bulk transfers.
This keeps the host CPU off the hot path; the host only launches jobs and observes results.

\textbf{Control plane.}
A small gRPC channel conveys mount/open/close, directory ops, and capability exchange (e.g., memory registration handles, QoS tokens).
Control messages are few and latency-insensitive relative to bulk I/O.

\textbf{Data plane.}
The DPU registers large receive/send buffers and drives the transport:
\begin{itemize}
  \item \textbf{TCP}: \texttt{ofi+tcp;ofi\_rxm} (OFI/libfabric) or \texttt{ucx+tcp} (UCX).
  \item \textbf{RDMA}: \texttt{ucx+rc,ucx+dc\_x} (UCX IB/RoCE) or \texttt{ofi+verbs;} \texttt{ofi\_rxm} (OFI/libfabric verbs).
\end{itemize}
Sequential I/O uses rendezvous-style transfers to amortize per-message overhead; random I/O uses short transfers but preserves zero-copy where possible.
All payloads currently terminate in DPU DRAM; the DPU notifies completion to the caller (FIO/DFS).

\subsection{Object Storage Server: DAOS I/O Engine}
\textbf{Network fabrics.}
DAOS’s RPC stack (CaRT/Mercury \cite{mercury}) runs over UCX or libfabric.
Each I/O engine is configured with a single fabric provider (e.g., \texttt{ucx+tcp}, \texttt{ucx+dc\_x}), and clients use a matching provider.
The DAOS I/O engine executes entirely in user space with kernel-bypass I/O—SPDK for NVMe and PMDK for SCM; UCX/libfabric for networking.
DFS is a client-side library that maps a POSIX-like namespace onto DAOS containers; servers export pools/containers and service the resulting object/metadata RPCs.

\textbf{Storage tiers.}
The engine accesses persistent tiers through PMDK (SCM) and SPDK (NVMe) with direct I/O queues.
This avoids kernel block-layer overhead and lets the server scale with the number of SSDs.

\textbf{DFS mapping.}
The DFS layer maps POSIX files and directories to DAOS objects and metadata entries.
Read/Write/RandRead/RandWrite from FIO translate into aligned object I/O (extents), with client-side batching for large requests and per-I/O checks for durability when requested.

\subsection{Protocol Choices and Hardware Interfaces}
Our design makes protocol–device mappings explicit:
\begin{itemize}
  \item \textbf{DPU}\,$\leftrightarrow$\,\textbf{DAOS server}: UCX or libfabric over TCP or RDMA (data plane).
  \item \textbf{DAOS server}\,$\leftrightarrow$\,\textbf{NVMe}: SPDK queues pinned to storage cores.
  \item \textbf{DAOS server}\,$\leftrightarrow$\,\textbf{SCM}: PMDK for low-latency, byte-addressable access.
  \item \textbf{Host}\,$\leftrightarrow$\,\textbf{DPU}: gRPC control channel; no payload bytes traverse the host kernel in the fast path.
\end{itemize}
This separation lets us compare the network provider (\texttt{ofi+tcp} vs.\ \texttt{ucx+tcp} vs.\ RDMA) without changing the application or the storage engine.

\subsection{Optional GPU Placement via GPUDirect RDMA}
While our prototype terminates data in DPU DRAM, the same architecture naturally supports optional GPU placement:
\begin{enumerate}
  \item The application (or a host helper) registers GPU buffers; the runtime obtains MR keys via \texttt{nvidia-peermem}.
  \item The control plane conveys the GPU buffer descriptors (addresses, sizes, rkeys) to the DPU and then to the storage server.
  \item On reads, the storage server performs RDMA writes that target the GPU buffer; on writes, the DPU (or server) sources data directly from the registered GPU memory.
\end{enumerate}
This \emph{GPUDirect RDMA} path keeps the same control/data-plane split and requires no changes to the DAOS engine; it simply replaces the DPU-DRAM sink/source with GPU HBM when the hardware topology allows it.
In this paper we report results with TCP and RDMA where data lands in DPU memory, and leave GPU placement as incremental work.

%% file: evaluation.tex
\section{Evaluation}

We evaluate \textsc{ROS2} to answer these questions:
\begin{itemize}
  \item How much overhead does DAOS (DFS) over TCP introduce relative to raw NVMe SSD ceilings in local and remote environments?
  \item What performance gains does RDMA provide over TCP for POSIX-compatible DAOS on server-grade CPUs?
  \item When offloading the DAOS client to an NVIDIA BlueField-3 DPU, how do TCP and RDMA perform relative to the host?
\end{itemize}

\subsection{Hardware Platform}

The storage server is equipped with 2 NUMA nodes with 128 CPU cores and 251GiB memory in total.
We pin the experiment to use NUMA node 0, which is connected with 4 NVMe SSDs (6.4TB in total) and ConnectX-6 (up to 200Gbps per port).

For the client, we do experiments in server-grade CPU and Bluefield-3. 
The CPU platform is a dual-socket server featuring two AMD EPYC 7443 24-core processors (48 physical cores, 96 threads total), 251GiB of DRAM, and a 200Gbps Mellanox ConnectX-6 adapter.
The NVIDIA BlueField-3 DPU features a 16-core Arm Cortex-A78AE processor, 30GiB of onboard DRAM, and an integrated Mellanox ConnectX-7 network controller capable of up to 400Gbps throughput

The storage server and client are connected via 100Gbps switch which constrains the maximum throughput especially when multiple SSDs are enabled in the server.

\subsection{Local Benchmark of NVMe SSDs}\label{sec:local-nvme}

\begin{figure}[t]
  \centering
  \captionsetup[figure]{aboveskip=0pt, belowskip=0pt}
  \captionsetup[subfigure]{aboveskip=-1pt, belowskip=-1pt}
  \begin{subfigure}{0.48\textwidth}
    \centering
    \includegraphics[width=\linewidth]{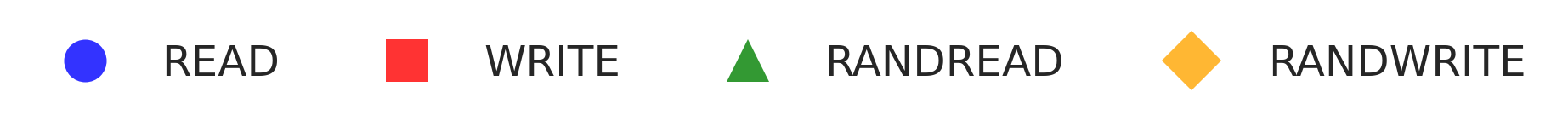}
  \end{subfigure}\hfill
  \begin{subfigure}{0.48\textwidth}
    \centering
    \includegraphics[width=\linewidth]{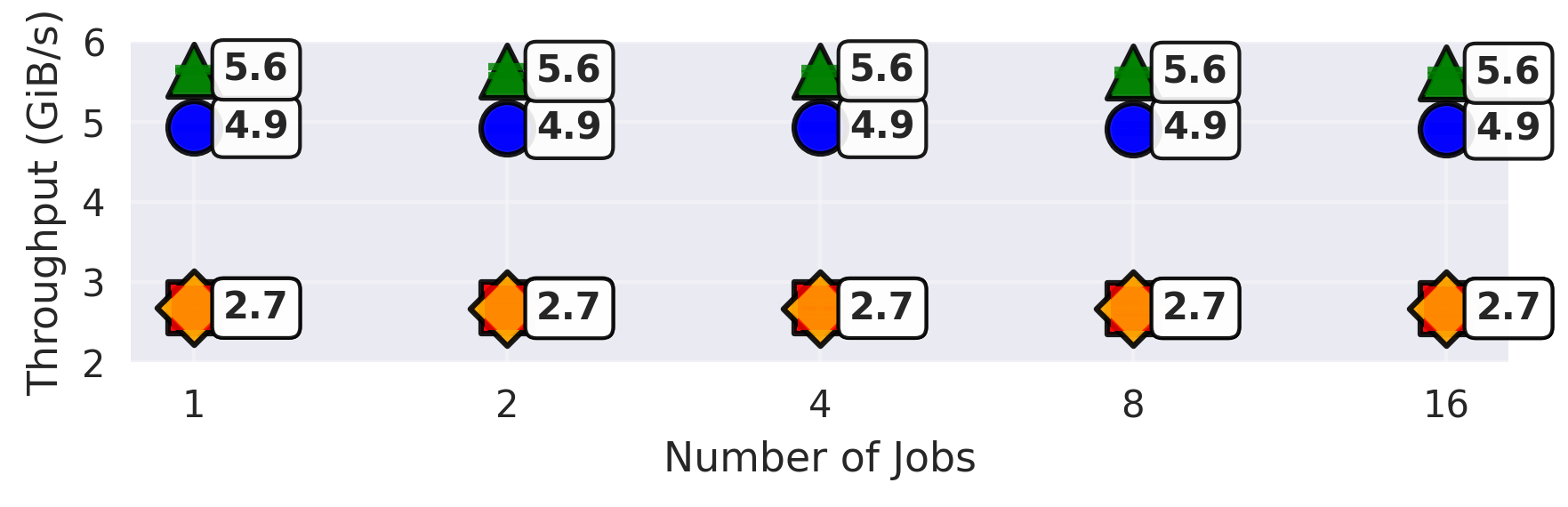}
    \caption{Throughput of block size 1MB for 1 NVMe SSD.}
    \label{fig:io_uring_1MB_1SSD}
  \end{subfigure}

  \begin{subfigure}{0.48\textwidth}
    \centering
    \includegraphics[width=\linewidth]{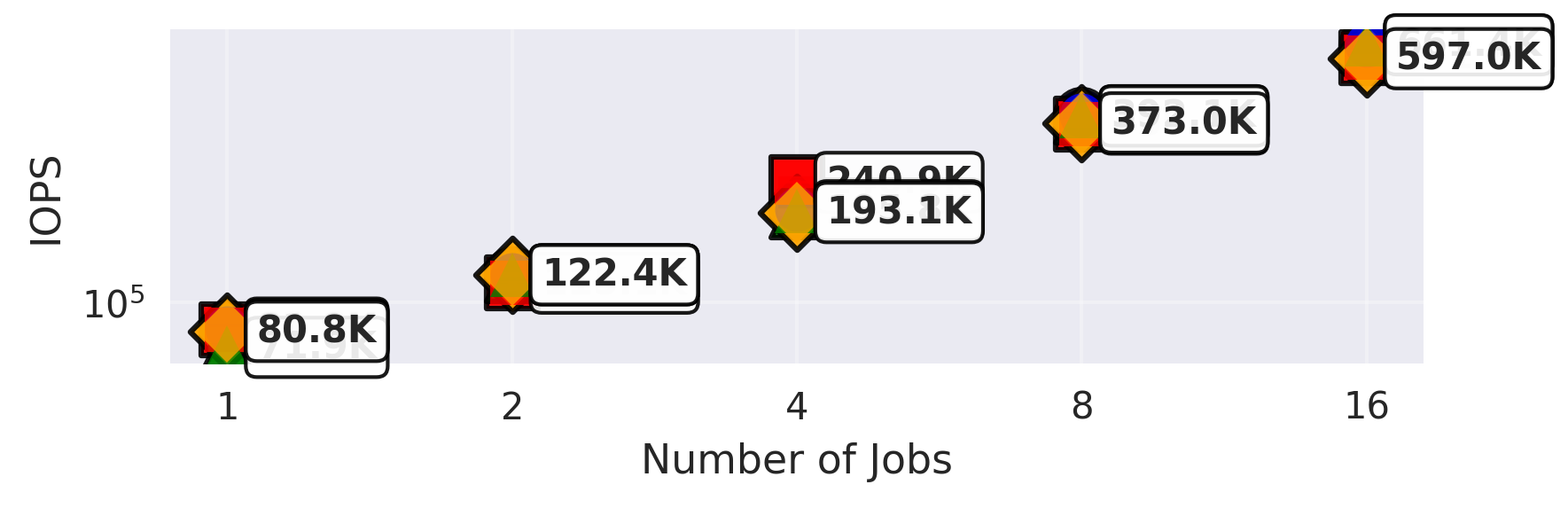}
    \caption{IOPS of block size 4KB for 1 NVMe SSD.}
    \label{fig:io_uring_4KB_1SSD}
  \end{subfigure}

  \begin{subfigure}{0.48\textwidth}
    \centering
    \includegraphics[width=\linewidth]{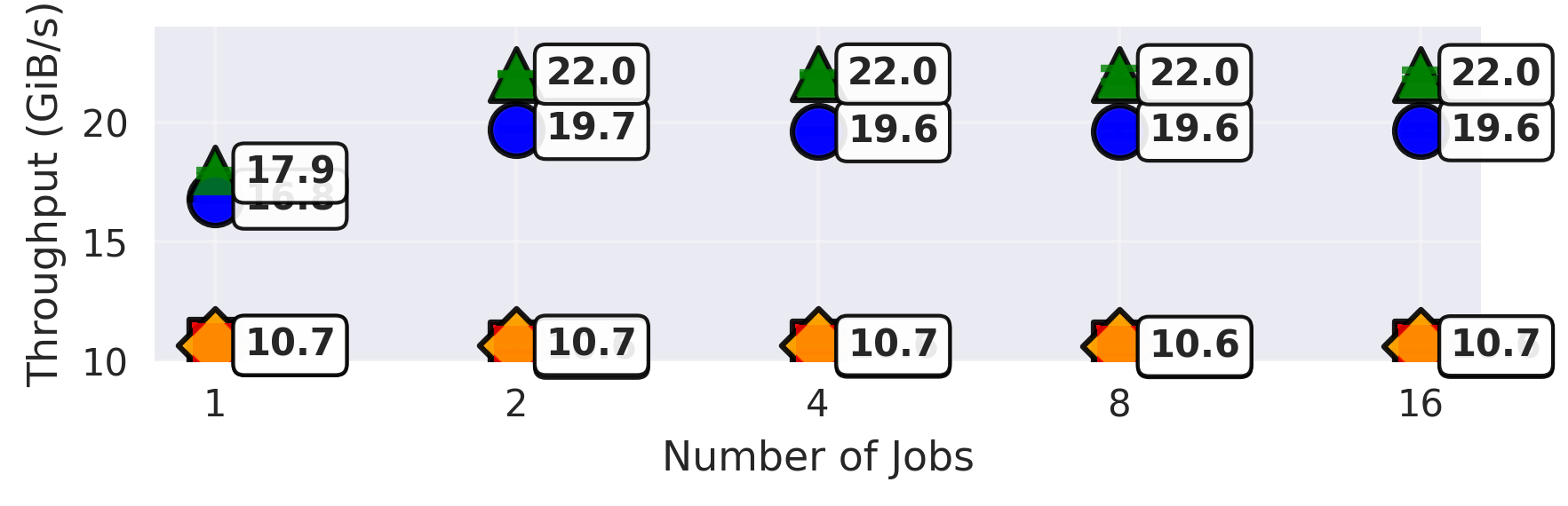}
    \caption{Throughput of block size 1MB for 4 NVMe SSDs.}
    \label{fig:io_uring_1MB_1SSD}
  \end{subfigure}

  \begin{subfigure}{0.48\textwidth}
    \centering
    \includegraphics[width=\linewidth]{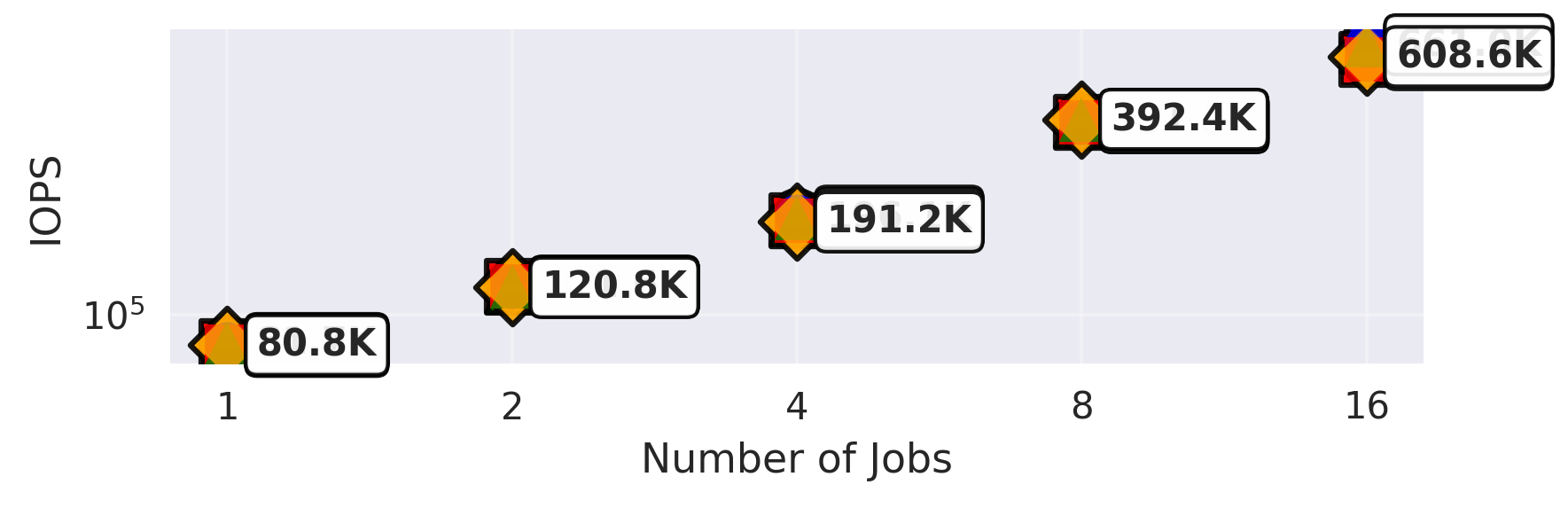}
    \caption{IOPS of block size 4KB for 4 NVMe SSDs.}
    \label{fig:io_uring_4KB_4SSD}
  \end{subfigure}
  \caption{Local FIO benchmark with IO\_URING engine with diverse number of SSDs.}
  \label{fig:local-fio}
\end{figure}

\noindent\textbf{Method.}
To establish device ceilings independent of the storage stack, we run FIO with the IO\_URING engine on a single node and vary the number of jobs (\{1,2,4,8,16\}) over two block sizes: 1\,MiB (throughput) and 4\,KiB (IOPS).
We report four POSIX-style workloads (read, write, randread, randwrite) for configurations with 1 and 4 NVMe SSDs (Fig.~\ref{fig:local-fio}).

\textbf{Results.}
(i) \emph{Large-block throughput saturates per device and scales with drive count.}
With 1\,SSD, sequential and random reads plateau around $\approx$5--5.6\,GiB/s, while writes plateau around $\approx$2.7\,GiB/s, and additional jobs provide no gain (Fig.~\ref{fig:local-fio}a).
With 4\,SSDs, throughput rises nearly linearly: reads reach $\approx$20--22\,GiB/s and writes $\approx$10.6--10.7\,GiB/s (Fig.~\ref{fig:local-fio}c).
(ii) \emph{Small-block IOPS require concurrency but show limited scaling with drive count.}
For 4\,KiB operations, IOPS grow with the number of jobs from $\sim$80\,K (1 job) to $\sim$600\,K (16 jobs) for both 1- and 4-SSD setups (Figs.~\ref{fig:local-fio}b,d), indicating a software/host-path limit rather than a single-drive media limit at low concurrency.
(iii) \emph{Random vs.\ sequential.}
At 1\,MiB, randread/randwrite closely track their sequential counterparts because transfer size dominates seek overhead; at 4\,KiB, the reverse holds, and access pattern plus submission concurrency determine IOPS.

\textbf{Implications.}
These measurements provide the local, stack-free baselines used later to normalize TCP and RDMA results.
They show that (a) one job suffices to saturate large-block per-device bandwidth, (b) adding drives yields near-linear gains for large transfers, and (c) achieving high small-block performance hinges on parallel submission rather than drive count alone.

\subsection{Remote Benchmark via SPDK}

\begin{figure*}[t]
  \centering

  \begin{subfigure}[t]{0.49\linewidth}
    \centering
    \includegraphics[width=\linewidth]{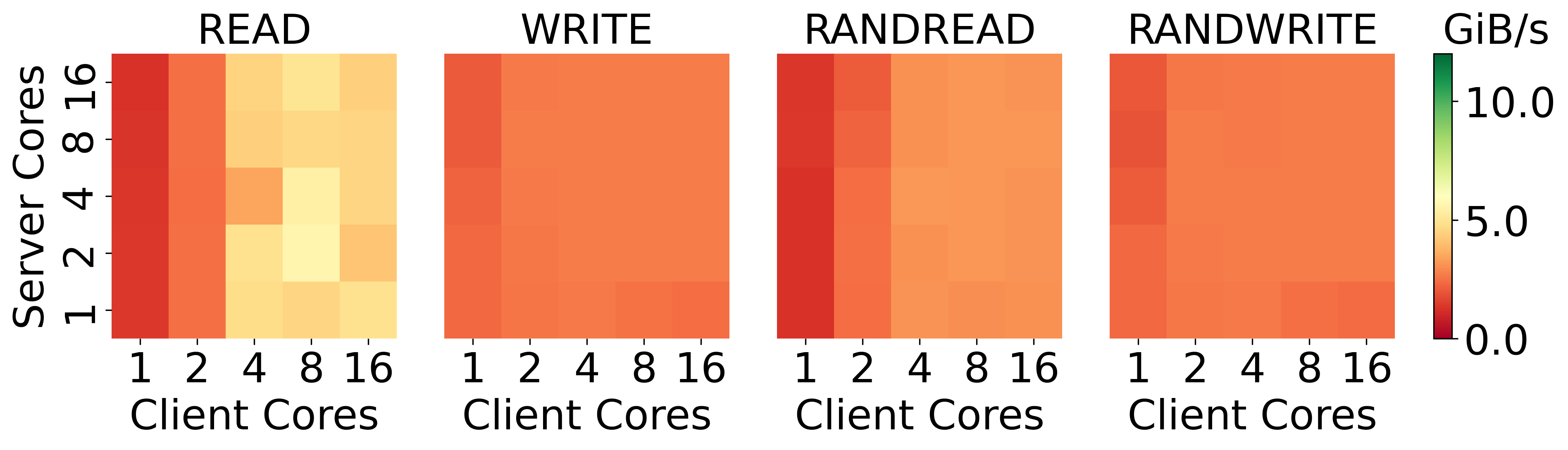}
    \caption{Throughput (1 MiB), TCP.}
    \label{fig:spdk_tcp_1M}
  \end{subfigure}\hfill
  \begin{subfigure}[t]{0.49\linewidth}
    \centering
    \includegraphics[width=\linewidth]{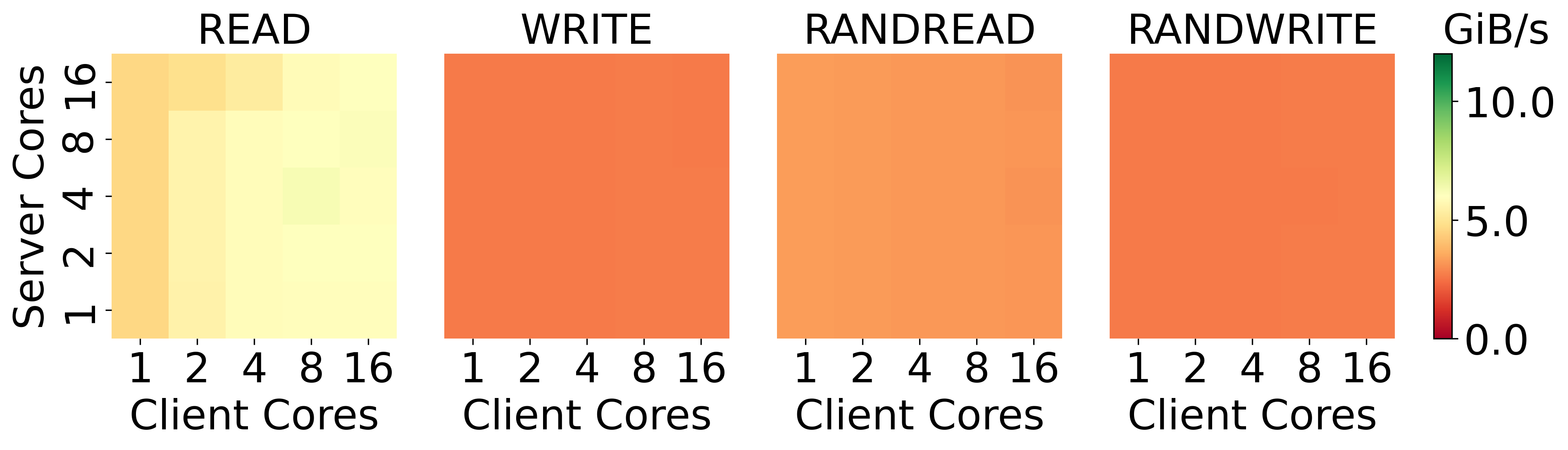}
    \caption{Throughput (1 MiB), RDMA.}
    \label{fig:spdk_rdma_1M}
  \end{subfigure}

  \vspace{2pt} 

  \begin{subfigure}[t]{0.49\linewidth}
    \centering
    \includegraphics[width=\linewidth]{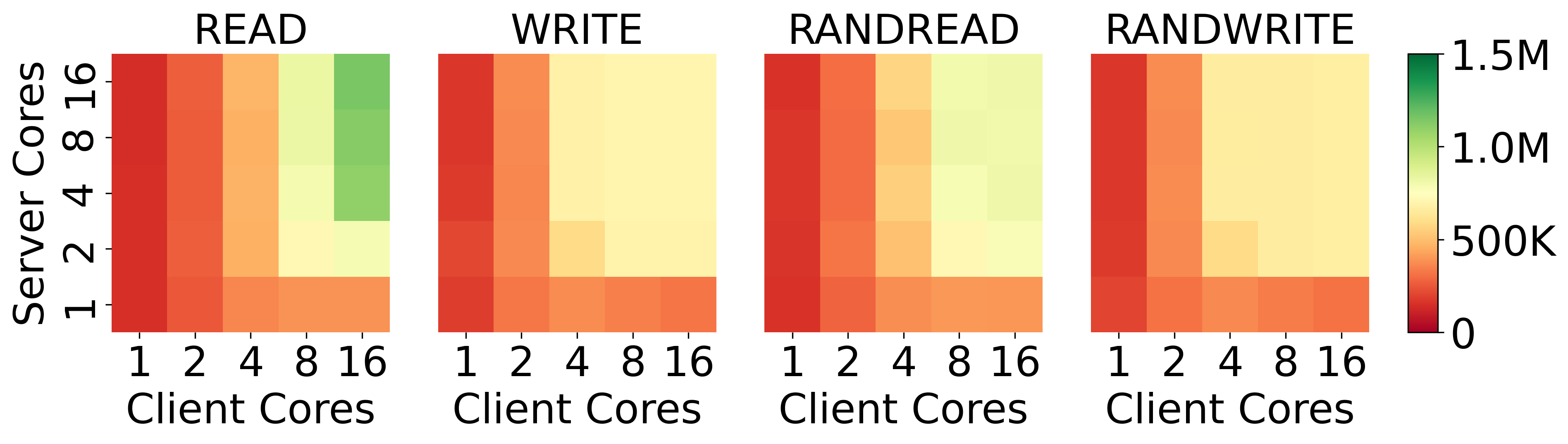}
    \caption{IOPS (4 KiB), TCP.}
    \label{fig:spdk_tcp_4K}
  \end{subfigure}\hfill
  \begin{subfigure}[t]{0.49\linewidth}
    \centering
    \includegraphics[width=\linewidth]{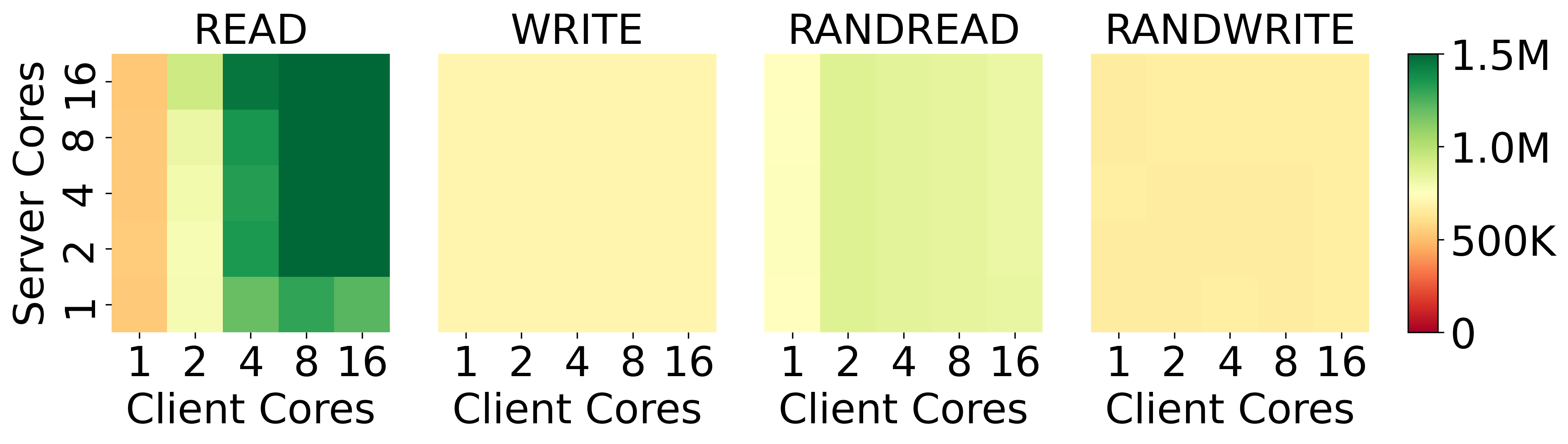}
    \caption{IOPS (4 KiB), RDMA.}
    \label{fig:spdk_rdma_4K}
  \end{subfigure}

  \caption{Remote SPDK benchmark of TCP and RDMA with 1 NVMe SSD.}
  \label{fig:spdk_benchmark}
\end{figure*}

\textbf{Method.}
We expose a single NVMe SSD from a storage node using the SPDK NVMe-oF target and drive it remotely from a client.
The client issues POSIX-style workloads (read, write, randread, randwrite) while varying \emph{client} and \emph{server} core counts (\{1,2,4,8,16\} on each side).
We compare transports: TCP and RDMA (Fig.~\ref{fig:spdk_benchmark}), and report large-block throughput (1\,MiB; Figs.~\ref{fig:spdk_tcp_1M}, \ref{fig:spdk_rdma_1M}) and small-block IOPS (4\,KiB; Figs.~\ref{fig:spdk_tcp_4K}, \ref{fig:spdk_rdma_4K}).

\textbf{Results.}
\emph{Large-block transfers (1\,MiB).}
Throughput improves with modest CPU parallelism on both ends but quickly plateaus for both transports (Figs.~\ref{fig:spdk_tcp_1M}, \ref{fig:spdk_rdma_1M}).
The similarity between TCP and RDMA at 1\,MiB indicates a media/network ceiling with one SSD; beyond a few cores, transport overheads are amortized.

\emph{Small-block transfers (4\,KiB).}
RDMA delivers substantially higher IOPS and better scaling with cores than TCP (compare Figs.~\ref{fig:spdk_tcp_4K} and \ref{fig:spdk_rdma_4K}).
The TCP heatmaps show limited benefit from additional client/server cores, while RDMA continues to gain, especially for reads/randreads, reflecting lower per-I/O latency and reduced CPU work in the fast path.

\textbf{Implications.}
(1) For large, sequential I/O on a single drive, transport choice matters little once concurrency is sufficient; the bottleneck shifts to media or link rate.
(2) For latency-sensitive small I/O, RDMA clearly outperforms TCP and benefits from CPU scaling on both client and server.
These results motivate using RDMA for POSIX-compatible DAOS over the network, and reserving TCP primarily for environments where RDMA is unavailable.

\subsection{Offloading DAOS onto SmartNIC}
\label{sec:dfs-on-dpu}

\begin{figure*}[t]
  \centering

  \begin{subfigure}[t]{0.48\linewidth}
    \centering
    \includegraphics[width=\linewidth]{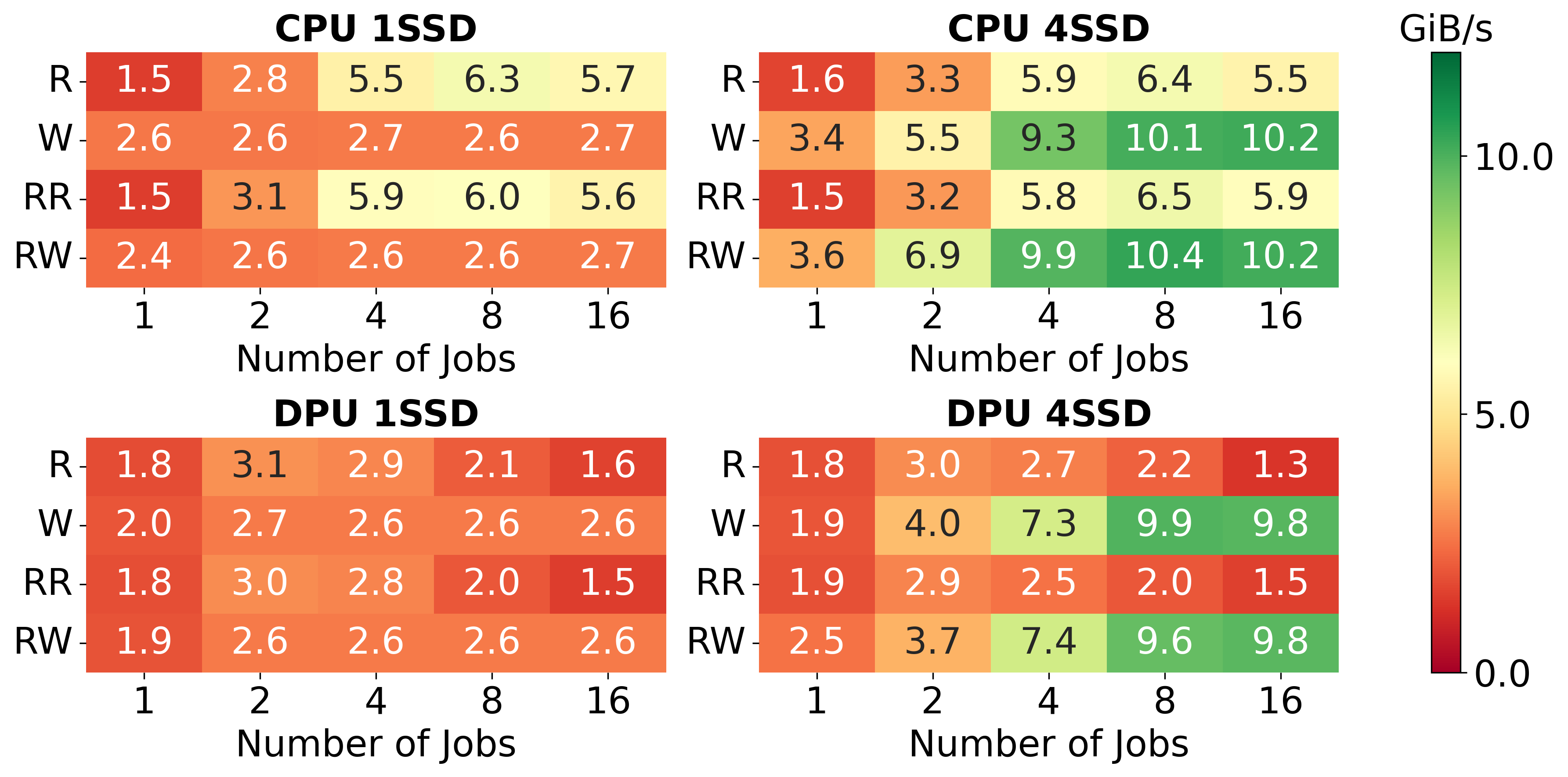}
    \caption{DFS TCP 1M}
    \label{fig:dfs_tcp_1M}
  \end{subfigure}
  \hfill
  \begin{subfigure}[t]{0.48\linewidth}
    \centering
    \includegraphics[width=\linewidth]{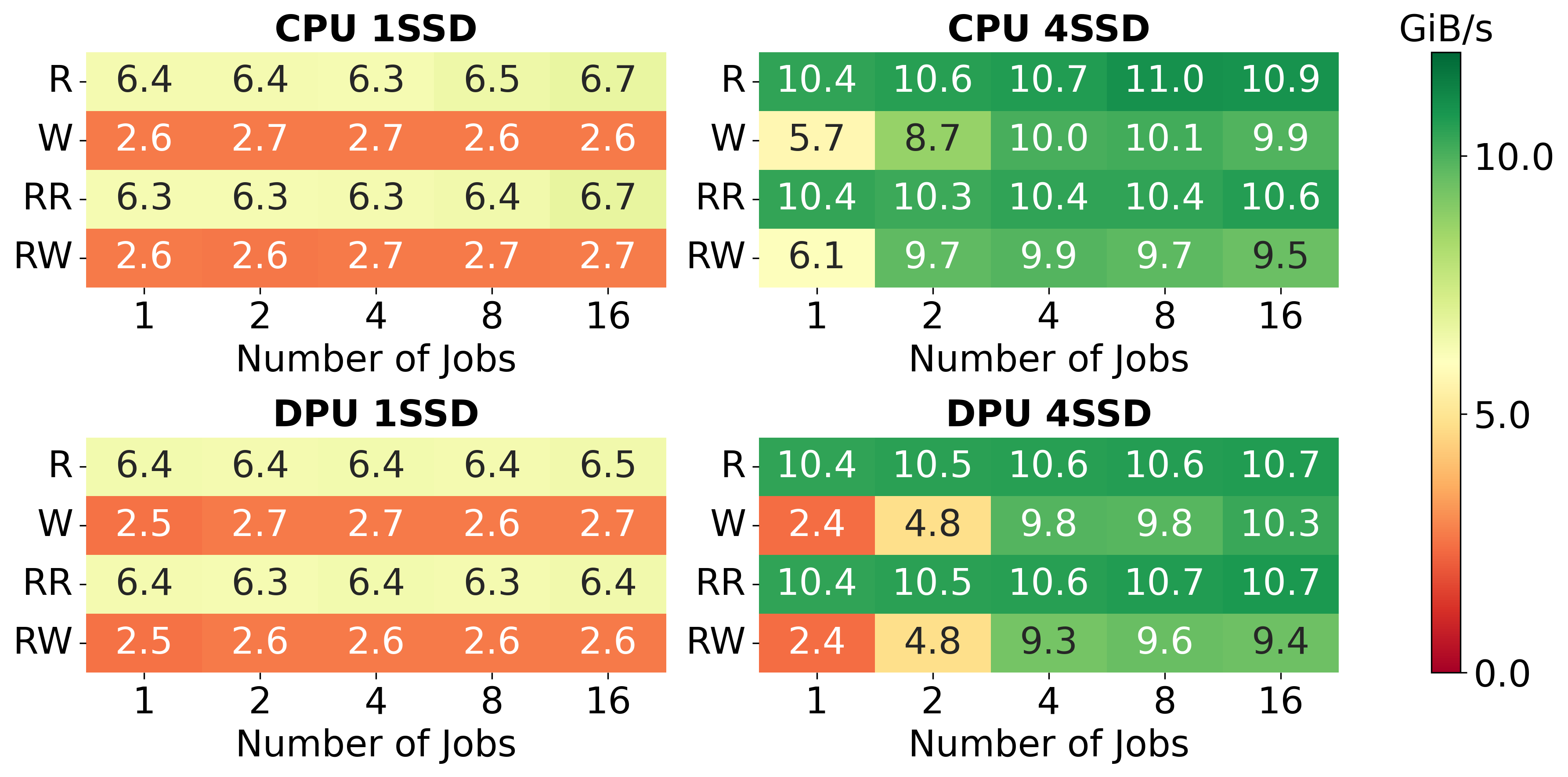}
    \caption{DFS RDMA 1M}
    \label{fig:dfs_rdma_1M}
  \end{subfigure}

  \par\smallskip 

  
  \begin{subfigure}[t]{0.48\linewidth}
    \centering
    \includegraphics[width=\linewidth]{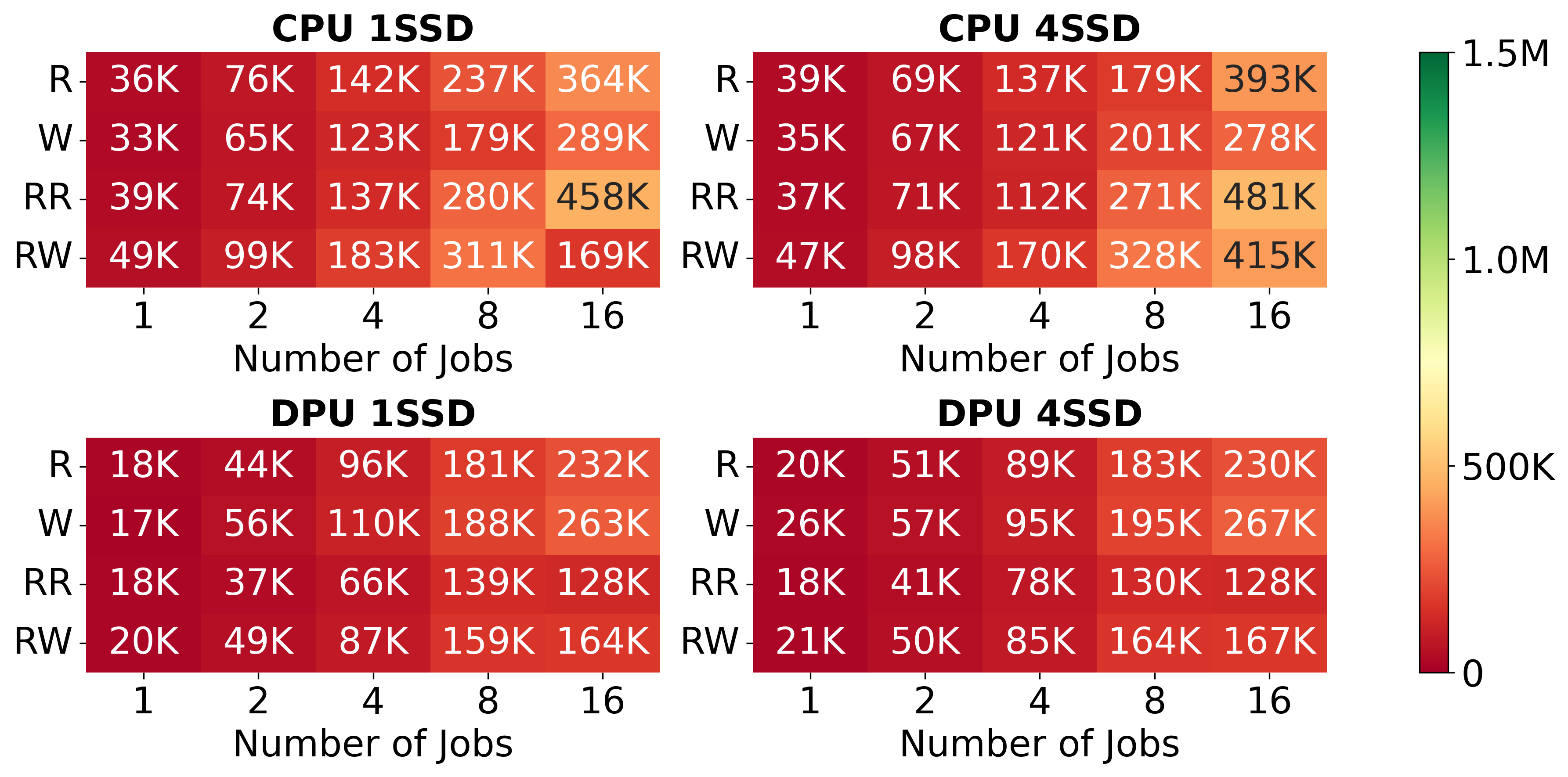}
    \caption{DFS TCP 4K}
    \label{fig:dfs_tcp_4K}
  \end{subfigure}
  \hfill
  \begin{subfigure}[t]{0.48\linewidth}
    \centering
    \includegraphics[width=\linewidth]{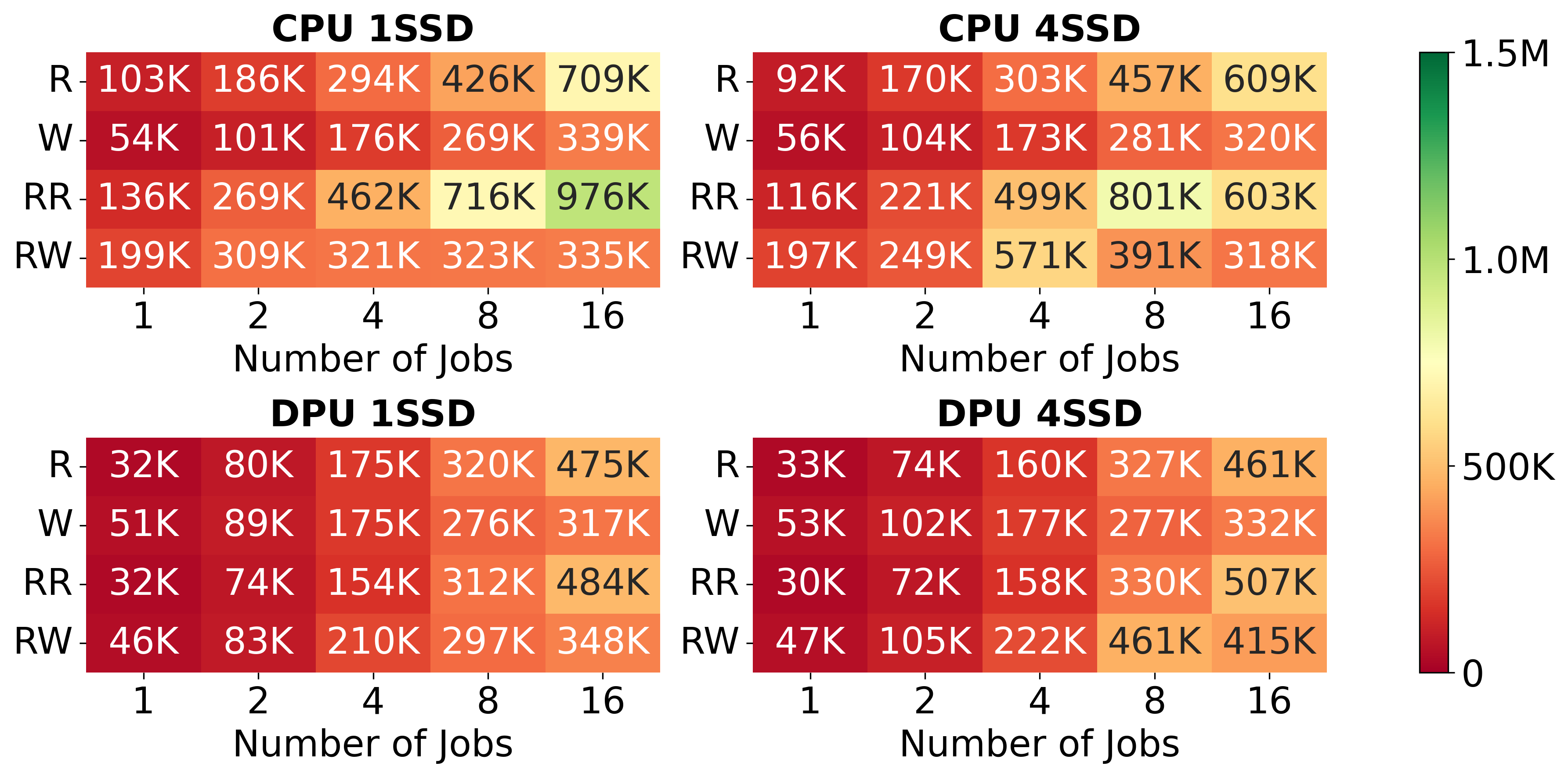}
    \caption{DFS RDMA 4K}
    \label{fig:dfs_rdma_4K}
  \end{subfigure}
  \caption{DFS results: TCP vs.\ RDMA. Left: 1\,MiB throughput (GiB/s).
        Right: 4\,KiB IOPS. Row labels use R = read, W = write, RR = random
        read, RW = random write.}
  \label{fig:dfs_grid}
\end{figure*}

\textbf{Setup.}
We offload the DAOS DFS client to an NVIDIA BlueField-3 (DPU) and compare it with a server-grade CPU host, using the same storage back end (1 vs.\ 4 NVMe SSDs) and the same workloads (read, write, randread, randwrite). We report large-block throughput (1\,MiB; Fig.~\ref{fig:dfs_tcp_1M}, \ref{fig:dfs_rdma_1M}) and small-block IOPS (4\,KiB; Fig.~\ref{fig:dfs_tcp_4K}, \ref{fig:dfs_rdma_4K}).

\textbf{TCP results.}
On the CPU, TCP reaches $\sim$5–6\,GiB/s with one SSD and $\sim$10\,GiB/s with four SSDs at 1\,MiB blocks (Fig.~\ref{fig:dfs_tcp_1M}, top), and scales to $\sim$0.4–0.6\,M IOPS at 4\,KiB (Fig.~\ref{fig:dfs_tcp_4K}, top). In contrast, the DPU’s TCP path underperforms: 1\,MiB \emph{reads} cap at $\sim$1.6–3.1\,GiB/s for one SSD and degrade with concurrency for four SSDs, while \emph{writes} with four SSDs can still approach $\sim$10\,GiB/s (Fig.~\ref{fig:dfs_tcp_1M}, bottom). For 4\,KiB I/O, the DPU tops out near $\sim$0.18–0.23\,M IOPS (Fig.~\ref{fig:dfs_tcp_4K}, bottom). The asymmetry (good TX, weak RX) indicates a DPU TCP receive-path bottleneck.

\textbf{RDMA results.}
RDMA removes the DPU penalty for large transfers: at 1\,MiB, the DPU matches the host for both one- and four-SSD setups ($\sim$6.4\,GiB/s and $\sim$10–11\,GiB/s, respectively; Fig.~\ref{fig:dfs_rdma_1M}). For 4\,KiB, RDMA on the DPU improves markedly over its TCP results (often 2$\times$ or more), though it still trails the CPU host by roughly 20–40\% depending on access pattern (Fig.~\ref{fig:dfs_rdma_4K}).

\textbf{Takeaways.}
(i) Offloading the DAOS client to BlueField-3 is \emph{performance-equivalent} to the host for large-block I/O when using RDMA.  
(ii) The DPU’s TCP path is unsuitable for read-heavy or small-block workloads, whereas RDMA narrows or eliminates the gap.  
(iii) For POSIX-compatible DFS over the network, RDMA is the preferred transport when deploying on SmartNICs.





%% file: discussion.tex
\section{Discussion}

Across SPDK (remote) and DAOS/DFS (end-to-end) experiments, RDMA consistently outperforms TCP for latency-sensitive, small-block workloads and matches TCP for large sequential transfers where media or link limits dominate. This reflects RDMA’s kernel-bypass and zero-copy data path, which avoid per-packet TCP processing and reduce host cycles per I/O.

Moving the DAOS client to BlueField-3 preserves RDMA performance—nearly identical to the server-grade CPU—while TCP on the DPU underperforms, indicating a receive-path bottleneck in the DPU TCP stack. In practice, this argues for an RDMA-first deployment when using DPUs: the offload still delivers isolation and multi-tenant control (dedicated QPs/PDs, per-tenant queues and rate limits) while keeping the host outside the hot path.

OS2G \cite{jin2025os2g} offloads a TCP/HTTP object-storage client to a DPU and adds a direct DPU\,$\rightarrow$\,GPU path (GDD) to reduce host involvement. Our results complement that direction but point to a different design point: when the storage stack natively supports RDMA (e.g., DAOS/DFS), an RDMA-driven object-storage path on BlueField attains CPU-class performance without paying TCP/HTTP overheads, making RDMA the preferable substrate for POSIX-compatible object I/O amid the growth of AI workloads.

For GPU-centric training clusters, a practical recipe is: expose storage tiers via userspace SPDK/PMDK; run DAOS/DFS over RDMA; and offload the client to DPUs for isolation and host-resource relief. Optional GPUDirect RDMA can then be layered to place data directly in GPU memory without revisiting transport or server design, yielding a minimal-copy data path.

Our study does not yet quantify host-side resource savings and does not integrate GPUDirect RDMA; both are promising follow-ups. We also plan to broaden device counts and NIC generations to stress multi-tenant scheduling and fairness on the DPU.

%% file: related_work.tex
\section{Related Work}

\textbf{RDMA in modern data centers.}
A growing body of work removes TCP/IP processing from storage data paths by adopting RDMA. 
DeepSeek 3FS \cite{deepseek3fs} demonstrates an RDMA-centric storage design tailored to LLM training that couples fast NVMe media with an RDMA transport and GPU-aware data movement. 
MinIO \cite{minio_s3_rdma} reports an S3-over-RDMA path that maps the S3 protocol onto RDMA to reduce latency and host CPU consumption, particularly for large objects and sustained throughput. 
At cloud scale, Azure \cite{bai2023empowering} details an end-to-end RDMA deployment across storage \emph{frontend} (compute$\leftrightarrow$storage) and \emph{backend} (intra-cluster) traffic, reporting broad regional rollout and significant CPU savings.

\textbf{SmartNIC offloading.}
Recent work shows that SmartNIC/DPUs, particularly NVIDIA BlueField series, substantially reduce host CPU usage, cut latency, and enable fair multi-tenant sharing by offloading data movement and control \cite{khalilov2024osmosis,khalilov2024network,xiao2024conspirator}. 
We extend this trend to RDMA-driven object storage I/O by offloading DAOS’s POSIX-compatible networking and storage components to BlueField, preserving throughput and lowering host resource consumption.

%% file: conclusion.tex
\section{Conclusion}
We presented an empirical study of POSIX-compatible DAOS for AI-centric workloads along media, transport, and offload dimensions. First, by characterizing raw NVMe SSD performance in both local and remote settings, we showed that device ceilings are readily attainable, indicating that storage media bandwidth is not the dominant bottleneck under our configurations. Second, on server-grade CPUs, benchmarking DAOS (DFS) over TCP and RDMA revealed that RDMA consistently outperforms TCP across sequential and random I/O, validating the benefits of kernel-bypass and zero-copy data paths. Third, when offloading the DAOS client to an NVIDIA BlueField-3 DPU, we observed markedly lower TCP performance than the x86 host, whereas RDMA on BlueField-3 matched host performance while shifting work off the host. Taken together, these results indicate that an RDMA-driven object storage stack with SmartNIC offload sustains high throughput and materially reduces host resource consumption, making it a compelling design point for large-scale AI data pipelines.